\documentclass[a4paper,11pt]{article}
\usepackage{fancyheadings}
\usepackage[dvips]{graphicx}
\usepackage{ifthen}
\usepackage{slashed,cite}
\usepackage{amssymb}
\usepackage{amsmath}
\usepackage{epsfig}
\usepackage[errorshow]{tracefnt}
\def\be{\begin{equation}}
\def\ee{\end{equation}}
\def\bea{\begin{eqnarray}}
\def\eea{\end{eqnarray}}
\def\nn{\nonumber}

\csname @addtoreset\endcsname{equation}{section}
\newcommand{\ft}[2]{{\textstyle\frac{#1}{#2}}}

\newsavebox{\uuunit}
\sbox{\uuunit}
    {\setlength{\unitlength}{0.825em}
     \begin{picture}(0.6,0.7)
        \thinlines
        \put(0,0){\line(1,0){0.5}}
        \put(0.15,0){\line(0,1){0.7}}
        \put(0.35,0){\line(0,1){0.8}}
       \multiput(0.3,0.8)(-0.04,-0.02){12}{\rule{0.5pt}{0.5pt}}
     \end {picture}}

\def\x{\xi}

\newcommand\benu{\begin{enumerate}}
\newcommand\eenu{\end{enumerate}}
\newcommand\bit{\begin{itemize}}
\newcommand\eit{\end{itemize}}

\newcommand\RR{{\mathbb R}}

\newcommand\wdg{\wedge}
\newcommand\del{\partial}
\newcommand\h{\frac{1}{2}}

\def\x{\times}

\begin{document}

\begin{flushright}
\small
UG-03-02 \\
{\bf hep-th/0303253}\\
\date \\
\normalsize
\end{flushright}

\begin{center}


\vspace{.7cm}

  {\LARGE {\bf Domain Walls}}

\vspace{.4cm}

   {\bf {\LARGE  and the}}

\vspace{.4cm}

    {\LARGE {\bf Creation of Strings}} \\

\vspace{1.2cm}

{\large Eric Bergshoeff,}
{\large Ulf Gran,}
{\large Rom\'an Linares,}

{\large Mikkel Nielsen}
{\large and}
{\large Diederik Roest}
\vskip 1truecm

\small
{\it Centre for Theoretical Physics, University of Groningen,\\
   Nijenborgh 4, 9747 AG Groningen, The Netherlands. \\ ~ \\
   E-mail: \tt (e.bergshoeff, gran, linares, nielsen, d.roest)@phys.rug.nl}

\vspace{.7cm}


{\bf Abstract}

\end{center}

\begin{quotation}

\small

The phenomenon of creation of strings, occurring when particles pass 
through a domain wall and 
related to the Hanany-Witten effect via dualities, is discussed in ten and 
nine dimensions.
We consider both the particle actions in massive backgrounds as well
as the 1/4-supersymmetric particle-string-domain wall supergravity solutions 
and discuss their 
physical interpretation.
In 10D we discuss the D0-F1-D8 system in massive IIA theory while in 9D 
the $SL(2,\mathbb{R})$-generalisation is constructed. It consists of 
$(p,q)$-particles, $(r,s)$-strings and the double domain wall solution of the 
three different 9D gauged supergravities where a subgroup of $SL(2,\mathbb{R})$ is gauged.

\end{quotation}

\newpage

\pagestyle{plain}

\section{Introduction}

It is well-known that D-branes can be understood as hyperplanes on
which a fundamental string, or F-string, can end \cite{Polchinski:1995mt}. 
The endpoint of an F-string appears
as an electrically charged particle on the worldvolume of the
D-brane and the corresponding worldvolume degrees of freedom are
associated to the Born-Infeld vector. An exception to this generic phenomenon
is the D-particle, on which a single F-string can not end due to charge conservation 
\cite{Strominger:1996ac, Townsend:1997em}.
This can be understood from the D-particle worldline point of view since
the Born-Infeld vector carries no degrees of freedom.

The situation changes in the presence of a domain wall
in which case charge
conservation no longer forbids an F-string to end on a D-particle \cite{Polchinski:1996sm}.
In fact, when a D-particle crosses a D8-brane, 
a stretched fundamental string with endpoints on the D0- and D8-brane is created 
\cite{Danielsson:1997wq,Danielsson:1998gi,Bergman:1997gf}.
This process is, via duality, related to the Hanany-Witten effect in which 
a stretched D3-brane is created if a D5-brane crosses an 
NS5-brane \cite{Hanany:1997ie}. The intersecting
configuration for this case 
is given by{}\footnote{Each horizontal entry indicates one of the 
10 directions $0,1,\ldots,
9$ in spacetime.  A $\x (-)$ means that the corresponding direction
is in the worldvolume of (transverse to) the brane.}
\begin{equation}
\label{hw1}
 \begin{array}{c|c}
{\rm D5}:\ \ \          \x & \x    \x   -   -   -  \x  \x   \x   - \\
{\rm NS5}:\          \x & \x   \x  \x  \x  \x  -  -   -   -    \\
{\rm D3}: \ \ \       \x & \x \x - - - - - - \x  
                         \end{array} 
\end{equation}
The intersecting configuration of  
\cite{Danielsson:1997wq,Danielsson:1998gi,Bergman:1997gf}
is obtained by
first applying T-duality in the directions 1 and 2, next applying an
S-duality and, finally, applying a T-duality in the directions 6,7 and
8:
\begin{equation}
\label{hw2}
 \begin{array}{c|c}
{\rm D0}:\ \ \          \x & -    -   -   -   -  -  - -    - \\
{\rm D8}:\ \ \          \x & \x   \x  \x  \x  \x  \x \x \x   -    \\
{\rm F1}: \ \ \       \x & - - - - - - - - \x  
                         \end{array} 
\end{equation}

In this paper we consider two ways to represent the process that a D-particle passing through a
domain wall leads to the creation of an F-string. First of all,
the creation of the F-string 
is represented by a coupling to the Born-Infeld 1-form
in the massive D0-brane action \cite{Bergshoeff:1996ui}.
Using an Einstein frame metric this action is given by (we use hats to indicate
that the corresponding object or index is ten-dimensional, after dimensional reduction we will omit the hats):
\begin{equation}
\label{term1}
S_{\rm massive\  D0-brane}
=T\int d\tau\,\Big(e^{-3\hat{\phi}/4}\sqrt{-\dot{X}^{\hat{\mu}}\dot{X}^{\hat{\nu}} \hat{g}_{\hat{\mu}\hat{\nu}}}+\dot{X}^{\hat{\mu}} \hat{A}_{\hat{\mu}}+m\,\hat{V}_t\Big)\,,
\end{equation}
where $T$ is the tension, $ {X}^{\hat\mu}(\tau)$ are
the embedding coordinates of the D-particle and $\hat{\phi}$ and 
$\hat{A}_{\hat{\mu}}$ are the IIA dilaton and R-R 1-form, respectively.
The Born-Infeld vector $\hat{V}_t$ appearing in the last term 
describes the tension of a fundamental string 
and the mass parameter $m$ signals the presence of a domain wall. 
Na\"{\i}vely, the field equation for $\hat{V}_t$ would imply a vanishing mass parameter. However, the above action is not complete as it stands, since we have to include new degrees of freedom corresponding to strings stretched between the D-particle and the D8-brane \cite{Banks:1997zs,Bergman:1997gf,Bachas:1998kn}. Once this is done, it is exactly the last term in \eqref{term1} that leads to strings ending on the D-particle.

A second way to investigate the creation of the fundamental string 
is by considering the following
solution\footnote{
There is a similar solution in the literature \cite{Janssen:1999sa}.
We will not 
consider this solution here for reasons that will be discussed 
later in the introduction.} 
\cite{Massar:1999sb} to the equations of motion of the $D=10$ Romans' massive IIA
supergravity theory \cite{Romans:1986tz}:
\begin{align}
\widehat{ds}^2 &= -H^{1/8}h^{-13/8}dt^2+H^{9/8}h^{-5/8}dy^2+H^{1/8}h^{3/8}dx_8^2 \,, \notag \\
\hat{B}_{ty}&=h^{-1} \,, \qquad
\hat{A}_t=H h^{-1} \,, \qquad
e^{\hat \phi} =H^{-5/4}h^{1/4} \,,
\label{IIAsolution}
\end{align}
where the harmonic functions $H$ and $h$ are defined as
\begin{align}
H=c+my\,,\quad h=1+\frac{Q}{r^6} \,,
\end{align}
and the radial coordinate is given in terms of the coordinates longitudinal to the D8-brane, $r^2=x_1^2+\ldots+x_8^2$.
The solution is 1/4 BPS and the Killing spinor is annihilated by the following projectors
\begin{align}
  \hat{\Pi}_{\text{D0}} = \tfrac{1}{2} (1+\Gamma^{\underline{0}}\Gamma_{11}) \,, \qquad
  \hat{\Pi}_{\text{F1}} = \tfrac{1}{2} (1+\Gamma^{\underline{0y}}) \,, \qquad
  \hat{\Pi}_{\text{D8}} = \tfrac{1}{2} (1+\Gamma^{\underline{y}}) \,,
\end{align}
where any of the three projectors can be obtained from the other two. 
The solution is a harmonic superposition of two elements, which can be obtained by taking different limits:
\begin{itemize}
\item The limit $Q \rightarrow 0$ leads to the single D8-brane solution which preserves 1/2 supersymmetry.
\item The limit $m \rightarrow 0$ leads to an (infinite)
F-string with D-particles smeared in the string direction, preserving 
1/4 supersymmetry. The F1- and D0-brane charges are related and therefore 
it is not possible to obtain these as single objects from the 
above solution.
\end{itemize}
More precisely, the flux distributions of the F-string and D-particle described 
by the solution (\ref{IIAsolution}) are given by
\begin{align}\label{Q}
  \hat{\mathcal{Q}}_1 & = e^{-\hat{\phi}} \star (d \hat{B}) = Q H \, d \Omega_7 \,, \notag \\
  \hat{\mathcal{Q}}_0 & = e^{3\hat{\phi}/2} \star (d \hat{A} - m \hat{B}) = Q \, dy \wedge d \Omega_7 \,,
\end{align}
with $d\Omega_7$ the volume form of $S^7$. 
To obtain the corresponding charges these are to be integrated over $S^7$ and $S^7 \times \mathbb{R}$, respectively, where  $S^7$ together with the 8D radius $r$ spans $\mathbb{R}^8$  and $\mathbb{R}$ covers the $y$-direction transverse to the domain wall.
The flux distributions are related by 
\begin{align}
  d \hat{\mathcal{Q}}_1 & = m \hat{\mathcal{Q}}_0 \,,
\end{align}
as required by the field equation for $\hat{B}$. 
This relation shows that in the presence
of a domain wall ($m \ne 0$), the D-particle ($\hat{\mathcal{Q}}_0\ne 0$) leads to the
creation of a fundamental string ($ d \hat{\mathcal{Q}}_1\ne 0$).
A similar result was obtained in \cite{Imamura:2001cr} for the NS5-D6-D8 system, i.e.~when
a NS5-brane passes through a D8-brane a D6-brane, stretched between the
NS5-brane and the D8-brane, is created. Both processes are
related to the Hanany-Witten effect via duality.

We now turn to the physical interpretation of the particle-string-domain wall solution
(\ref{IIAsolution}). It is instructive to first discuss some general issues of domain walls.
To prevent the vanishing of the harmonic function $H$ of the domain wall, which would 
imply $e^{\hat{\phi}} \rightarrow \infty$, one has to add a source term in the 
transverse space, say at $y=0$. There are two distinct objects one can include:
\begin{itemize}
\item
An object with negative tension, corresponding to an O8-plane which 
arises by dividing out by $\mathbb{Z}_2$.
The resulting harmonic function will be 
$H = c + m |y|$ with $c$ and $m$ positive \cite{Chamblin:1999ea, Bergshoeff:2001pv}. 
Indeed $H$ is positive for all values of $y$ and has a minimum at the 
O8-plane.
\item
An object with positive tension, corresponding to a D8-brane. 
Passing through such a domain wall leads to a decrease of the slope of the harmonic function 
\cite{Polchinski:1995df, Chamblin:1999ea, Bergshoeff:2001pv}. The prime example is
$H = c - m |y|$ with $c$ and $m$ positive. It follows that $H$ will vanish for some 
critical value of $y$. 
\end{itemize}
One thus finds that the introduction of D8-branes leads to zeroes in $H$ and 
thus to a 'critical distance'.
It forces one to include O8-planes at a smaller distance, such that the zero in $H$
is avoided. In this case the relevant $\mathbb{Z}_2$
symmetry is given by $I_y\Omega$, where $I_y$ is a reflection
in the transverse space and $\Omega$ is the worldsheet parity operator.
If the transverse space is $\mathbb{R}/\mathbb{Z}_2$, we can
take one O8 plane with R-R charge -16 and $n$ D8-branes and their images with $n \le 8$.
For $n>8$ the total tension is positive and a zero in the harmonic will
occur.
On the other hand if the transverse space is $S^1/\mathbb{Z}_2$, i.e.
$y$ is compact, the total tension has to vanish and one is led to 
type I${}^\prime$ string theory with two O8 planes at the two fixed points and
16 D8-branes and their images in between \cite{Polchinski:1995df}.

We now return to the distribution of D-particles and F-strings described by 
(\ref{IIAsolution}). First of all we note that all non-zero tensor components
of (\ref{IIAsolution}) are even under the relevant $\mathbb{Z}_2$-symmetry $I_y \Omega$.
If this were not the case, see e.g. some of the solutions of \cite{Janssen:1999sa, Imamura:2001cr}, one is forced to include a source term, corresponding 
to the non-zero tensor components, which is smeared over a 9D hyperplane.
The only odd field that we allow for is the mass parameter and the 
corresponding source terms are the domain walls.
The supergravity solution (\ref{IIAsolution}) that we consider only has
even non-zero tensor components  
and the inclusion of source terms for this solution was discussed in 
\cite{Kallosh:2001zc}\footnote{The particle and strings source terms of 
\cite{Kallosh:2001zc} are smeared in the $y$-direction and directly relate 
to the charge distributions \eqref{Q}.}, resulting in a globally well-defined solution
on $S^1 / \mathbb{Z}_2$. We will now discuss its physical implications.

\begin{figure}[h]
  \centerline{\epsfig{file=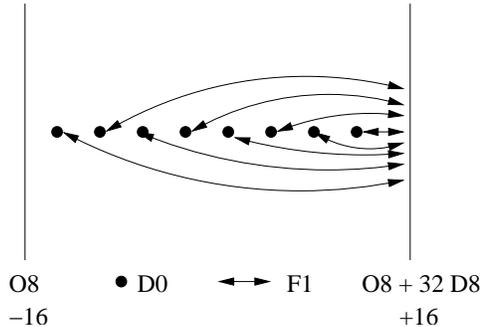,width=.45\textwidth}}
  \caption{\it The creation of strings in type I${}^\prime$: a (continuous) distribution of D-particles with
  a monotonically increasing distribution of unoriented F-strings ending on the D8-branes.
  The distribution of these F-strings has a maximum at the position of the D8-branes.} \label{HWDO}
\end{figure}

We note that the distribution of F-strings is linear 
in $H$, see eq.~(\ref{Q}). When we are dealing with a D8-brane, we have $H= c- m|y|$
which is a linearly increasing function when going towards the domain wall.
This is in agreement with the idea of creation of strings when passing 
through a D8-brane \cite{Danielsson:1997wq,Danielsson:1998gi,Bergman:1997gf}. 
It is pictorially given in figure \ref{HWDO}, where we have taken all D8-branes to 
coincide with one of the orientifolds. 
The strings are unoriented due to the identification $y \sim - y$ which superposes two strings
of opposite orientation, see e.g. \cite{Bergman:1997gf}.
It should be noted that the linear behaviour of $\mathcal{Q}_1$ is an artifact of the coordinate 
frame for the transverse coordinate $y$. The important feature is that it is monotonically increasing when approaching the domain wall. 

It is the aim of this paper to investigate the above situation and to generalise it to 
$D=9$ dimensions. We will show that in $D=9$ there is a natural 
$SL(2,\mathbb{R})$-generalisation which consists of 
$(p,q)$-particles, $(r,s)$-strings and the double domain wall 
solution \cite{Bergshoeff:2002mb} of the three different 9D gauged 
supergravities where a subgroup of $SL(2,\mathbb{R})$ is gauged
\cite{Meessen:1998qm,Hull:1998vy,Bergshoeff:2002mb,Hull:2002wg}. 
We will describe this situation both
from the point of view of a $(p,q)$-particle action in a massive background as well as
from the corresponding $D=9$ particle-string-domain wall solution.

The outline of the paper is as follows. In section \ref{9Dbrane}, we perform a
Scherk-Schwarz double dimensional reduction of the $(p,q)$-string in a type IIB
background and obtain the action for a $(p,q)$-particle
in a gauged 9D supergravity background. In section \ref{9Dsugra}, we 
construct the general $D=9$ particle-string-domain wall supergravity 
solution that corresponds to the creation of strings in 9D.
In both sections (actions and solutions) we discuss the relation 
between the new results in 9D and the known 
examples in 10D. Finally, in section \ref{discussion} we discuss our results,
their physical interpretation and conclude. We have included three Appendices.
Appendix A gives our conventions, appendix B gives some technical details
related to the Scherk-Schwarz double dimensional reduction and 
appendix C discusses different formulations of the $(p,q)$-string action.

\section{The $(p,q)$-Particle in Gauged 9D Supergravity}\label{9Dbrane}

In this section we will generalise the D-particle action \cite{Bergshoeff:1996cy} in
the massive IIA background to the $(p,q)$-particle action in the
different 9D gauged supergravity backgrounds of \cite{Meessen:1998qm}. 

\subsection{The $(p,q)$-String in IIB} \label{pqstring}

Our starting point is
the $SL(2,\RR)$-covariant action \cite{Townsend:1997kr,Cederwall:1997ts} 
(our notation can be found in appendix \ref{conventions})
\be\label{l}
S=\h\int d^2\sigma\lambda\,(\hat{g}+\Phi \bar{\Phi})\, ,
\ee
where 
\begin{align}
  \Phi= \h\epsilon^{ij}\hat{{\cal F}}_{ij}=\ast\hat{{\cal F}}\,, \qquad
  \hat{{\cal F}}= e^{\hat{\phi}/2} (\hat{\tau} \hat{F}_1 + \hat{F}_2) \,, \qquad
  \hat{F}_r=d\hat{V}_r-\hat{B}_r \,.
\end{align}
Here $\hat{V}_r$ is the doublet of vectors living on the worldvolume, 
with $r=1,2$ referring to the NS-NS and R-R sector, respectively. 
For the D-string, the doublet of vectors is the Born-Infeld vector as well as a vector generating the D-string tension. In order to apply the 
Scherk-Schwarz reduction in the following section, it is convenient
that we work with an $SL(2,\RR)$-covariant action. 
The complex scalar $\hat{\tau} = \hat{\chi} + i e^{-\hat{\phi}}$ consists of the
axion $\hat{\chi}$ and the dilaton $\hat{\phi}$ and $\hat{g}$ is the determinant of $\hat{g}_{ij}$, which is the Einstein frame metric.
The hats indicate that we are dealing with a string in a 10D 
background, whereas unhatted fields will be used after dimensional reduction.
The background fields are pulled back to the worldvolume, i.e.
\begin{align}
\hat{g}_{ij}=\del_i X^{\hat{\mu}}\del_j X^{\hat{\nu}}\hat{g}_{\hat{\mu}\hat{\nu}}\, \qquad
\hat{B}_{ij}=\del_i X^{\hat{\mu}}\del_j X^{\hat{\nu}}\hat{B}_{\hat{\mu}\hat{\nu}}\, .
\end{align}
We can write the last term in (\ref{l}) as
\be 
 \Phi \bar{\Phi} = \hat{{\cal M}}^{rs}\ast \hat{F}_r 
\ast \hat{F}_s\, ,
\ee
where the matrix $\hat{{\cal M}}$ can be written in terms of the axion and the
 dilaton as
\be\label{scal}
\hat{{\cal M}}=e^{\hat{\phi}}\,\left(\begin{array}{cc}\vert\hat{\tau}\vert^2&
\hat{\chi}\, ,\\
\hat{\chi}&1\end{array}\right)\, .
\ee
By using the field equation for the vectors one can dualise these into constants.
For instance, by integrating out both vectors, the action for the fundamental string is
obtained, whereas the D-string action follows after integrating out $\hat{V}_2$.
This is discussed in detail in appendix \ref{pqformulations}.

\subsection{Double Dimensional Reduction}

We will now perform a double dimensional Scherk-Schwarz reduction of the $SL(2,\RR)$-covariant string action in ten dimensions to obtain an $SL(2,\RR)$-covariant action of particles coupled to a massive supergravity background in nine dimensions \cite{Meessen:1998qm}. 
Unlike ordinary Kaluza-Klein reduction, where the background fields are assumed to be independent of the reduction coordinate, with Scherk-Schwarz reduction \cite{Scherk:1979ta} of type IIB supergravity the background fields get an $x$-dependence through an $SL(2,\RR)$-transformation, but in a way such that the reduced action is independent of the reduction coordinate. The reduction coordinate is now the spatial coordinate of the string, $x=\sigma$, and therefore we have to perform a Scherk-Schwarz reduction of both the background and the worldvolume fields. For the background, the $x$-dependent $SL(2,\RR)$-transformation is
\be
\Omega=\exp(xM)=\left(\begin{array}{cc}\cosh(\alpha x)+\frac{1}{2} m_1 \sinh(\alpha x)&\frac{1}{2\alpha} (m_2+m_3) \sinh(\alpha x)\\
\frac{1}{2\alpha}(m_2-m_3)\sinh(\alpha x)&\cosh(\alpha x)-\frac{1}{2} m_1 \sinh(\alpha x)\end{array}\right) \,,
\ee
where the mass matrix is defined as
\be
M^r{}_s=\tfrac{1}{2}\left(\begin{array}{cc}m_1&m_2+m_3\\
m_2-m_3&-m_1\end{array}\right) \,,
\ee
and we have used the $SL(2,\RR)$-invariant quantity 
\be
\alpha^2 = \det \, (M) = \frac{1}{4}(m_1^2+m_2^2-m_3^2)\, .
\ee
This gives rise to three different conjugacy classes of $SL(2,\RR)$, corresponding to $\alpha^2>0$, $\alpha^2=0$ and $\alpha^2<0$, which is just the light-cone structure of the isomorphic group SO(2,1) 
\cite{Hull:1998vy,Hull:2002wg, Bergshoeff:2002mb}. A representative for each class is given in table \ref{massparameters}.

\begin{table}[t]
\begin{center}
\begin{tabular}{||c||c|c|c||}
\hline \rule[-3mm]{0mm}{8mm}
  Gauge group & $\alpha^2$ & $(m_1,m_2,m_3)$ & $(\mu_1,\mu_2)$ \\
\hline \hline \rule[-3mm]{0mm}{8mm}
  $\mathbb{R}$ & $0$ & $(0,m,m)$ & $(0,2m)$ \\
\hline \rule[-3mm]{0mm}{8mm} 
  $SO(2)$ & $-m^2$ & $(0,0,2m)$ & $(2m,2m)$ \\
\hline \rule[-3mm]{0mm}{8mm} 
  $SO(1,1)$ & $m^2$ & $(0,2m,0)$ & $(-2m,2m)$ \\
\hline
\end{tabular}
\caption{\it The different choices of mass parameters and the corresponding different gauge groups. For later reference we also give $\mu_1 = -m_2+m_3$ and $\mu_2=m_2+m_3$. \label{massparameters}}
\end{center}
\end{table}

The reduction Ansatz can be found in appendix B. For the fields that are tensor valued under $SL(2,\RR)$ we have
\begin{align}
\hat{{\cal M}}^{rs}=\Omega^r{}_p \, \Omega^s{}_q \,{\cal M}^{pq}\,,\quad 
\hat{B}_r=(\Omega^{-1T})_r{}^s B_s\, ,
\end{align}
where unhatted fields are 9D fields. 
Since the nine-dimensional coordinates $X^\mu$ are independent of $\sigma$, the 1-components of the 9D fields pulled back to the worldline are zero. The pulled-back quantities become
\begin{align}
\hat{g}_{ij}&=e^{-\sqrt{7}\varphi/2}\left(\begin{array}{cc}\dot{X}^\mu \dot{X}^\nu g_{\mu\nu} +\dot{X}^\mu \dot{X}^\nu A_\mu A_\nu&-\dot{X}^\mu A_\mu\\
-\dot{X}^\mu A_\mu&1\end{array}\right)\, ,\notag \\
(\hat{B}_r)_{01}&=-\dot{X}^\mu (A_r)_\mu\, .
\end{align}
The worldvolume vector should now be transformed with the $SL(2,\RR)$ matrix and therefore acquires a $\sigma$-dependence, in contrast to ordinary Kaluza-Klein reduction, where $\sigma$-derivatives of the worldvolume vector can be dropped. Because of the worldline indices, the doublet of vectors splits into a doublet of scalars and a doublet of worldline vectors
\bea
(\hat{V}_r)_1=(\Omega^{-1T})_r{}^s S_r\, ,\quad
(\hat{V}_r)_0=(\Omega^{-1T})_r{}^s V_r\, .
\eea
The resulting action is $\sigma$-independent and the integration over $\sigma$ therefore just yields a constant, which we can normalise to 1. The  action can be interpreted as the action for 0-branes with charges $q^r$ coupled to 9-dimensional massive supergravity
\be
S=\h\int d\tau\,\lambda\,\Big(e^{-3\varphi/\sqrt{7}}\dot{X}^\mu\dot{X}^\nu g_{\mu\nu}+{\cal M}^{rs}P_r P_s\Big)\, ,
\ee
where
\be
P_r=\dot{S}_r+\dot{X}^\mu (A_r)_\mu+M^s{}_r V_s\, .
\ee
The charges $q^r = (p,q)$ are encoded in the worldline scalars $S^r$ and refer to the NS-NS and R-R sector, respectively.
Indeed, one can manipulate the form of the action by dualising the 
scalars $S_r$ into constants $q^r$\footnote{
  This is analogous to the different forms of the $(p,q)$-string action, where one can dualise 
  the vectors $\hat{V}_r$ into constants $\hat{q}^r$. This is discussed in detail in appendix
  \ref{pqformulations}.}.
To make the charges explicit, we
 add a surface term without changing the equations of motion:
\be
S=\h\int d\tau\,\lambda\,\Big(e^{-3\varphi/\sqrt{7}}\dot{X}^\mu\dot{X}^\nu g_{\mu\nu}+{\cal M}^{rs}P_r P_s\Big)+q^r\int d\tau\,\dot{S}_r\, .
\ee
Using the equations of motion for $\lambda$ and the doublet of worldvolume scalars $\dot{S}_r$, we get an action which depends on the metric via a square root
\be
S=\int d\tau\, \sqrt{q_r {\cal M}^{rs} q_s} \, e^{-3\varphi/2\sqrt{7}}\sqrt{-\dot{X}^\mu\dot{X}^\nu g_{\mu\nu}}-q^r\int d\tau\,(\dot{X}^\mu(A_r)_\mu+M^s{}_r V_s)\, .
\label{particleaction}
\ee
It is the last term that is of particular interest to us and it is the natural generalisation 
of the 10D IIA coupling appearing in \eqref{term1}.
It is related to the generalisation of the creation of strings in a massive background since
\begin{itemize}
\item 
The $q^r$ are the NS-NS and R-R charges of the $(p,q)$-particle, respectively.
\item
The matrix $M^s{}_r$ characterises the massive background and specifies which of the three gauged supergravities it corresponds to.
\item
The $V_s$ are the worldline vectors that correspond to the endpoints of the 
D-string and F-string on the $(p,q)$-particle worldline, respectively.
\end{itemize} 
We will comment further on how the last term in this action is related
to the creation of strings in the discussion.
Before ending this section we show that the Kaluza-Klein reduction of the 
D-particle yields a special case of the actions constructed above.

\subsection{Relation to the D-Particle in Massive IIA}

Romans' massive type IIA supergravity in 10 dimensions is related to type IIB supergravity via massive T-duality. In particular, KK-reduction of Romans' supergravity to 9 dimensions yields the same result as Scherk-Schwarz reduction of type IIB with the mass parameters $m_1=0$ and $m_2=m_3=m$. Actions for D-branes coupled to Romans' massive supergravity have been obtained \cite{Bergshoeff:1996cy,Green:1996bh}. The result is a standard action plus a topological mass term. By direct KK-reduction of the massive D0-brane action, the particle action from the previous section should be reproduced with the above mass parameters. By applying the appropriate reductions, it was shown in \cite{Bergshoeff:1996cy} that the D0- and D1-brane actions yield the same action in nine dimensions. We will here show explicitly that the massive D0-brane action can be reduced to a special case of the $SL(2,\RR)$-covariant 0-brane action derived in the previous subsection.

The massive D0-brane action in 10 dimensions is given in \eqref{term1}.
As for the actions in the first subsections, 
there also exists a ``squared'' version of the D0-brane action, which is obtained by adding a tension field
\be
S=\int d\tau\,\tilde{\lambda}\,\Big(e^{-3\hat{\phi}/2}\dot{X}^{\hat{\mu}}\dot{X}^{\hat{\nu}} \hat{g}_{\hat{\mu}\hat{\nu}}+(\ast\tilde{F})^2\Big)-T\int d\hat{S}_2\, ,
\ee
where
\be
\tilde{F}=d\hat{S}_2+d\tau\,(\dot{X}^{\hat{\mu}} \hat{A}_{\hat{\mu}}+m\,\hat{V}_t)\, .
\ee
We now perform a KK-reduction of this action. The reduction coordinate becomes a worldvolume scalar $x=S_1$. Using the reduction Ansatz from appendix B, we get 
\bea\nn
&& S=\h\int d\tau\,\tilde{\lambda}\,\Big(e^{-3\varphi/\sqrt{7}}e^{-\phi}\dot{X}^\mu\dot{X}^\nu g_{\mu\nu}+(e^{-2\phi}+\chi^2)(\dot{X}^\mu (A_1)_\mu+\dot{S_1})^2\\
&&\qquad+2\chi(\dot{S_2}+\dot{X}^\mu (A_2)_\mu-m V_t)(\dot{S_1}+\dot{X}^\mu (A_1)_\mu)\Big)-T\int d\tau\, \dot{S_2}\, .
\eea
We can add a surface term for $\dot{S_1}$ without changing its field equation. 
Upon making the identifications $V_1=-\hat{V}_t$ the action looks as follows:
\be
S=\h\int d\tau\,\lambda\,\Big(e^{-3\varphi/\sqrt{7}}\dot{X}^\mu\dot{X}^\nu g_{\mu\nu}+{\cal M}^{rs}P_r P_s\Big)+q^r\int d\tau\,\dot{S}_r\, ,
\ee
where $q^2 = -T$ 
and we have used that the mass matrix in this case has the form
\be\label{mr}
M=\left(\begin{array}{cc}0&m\\0&0\end{array}\right)\, .
\ee
Finally, writing $q^1$ as an integer times the string tension, 
i.e. $q^1 = Tp$ with p integer, we conclude that
the massive D0-brane action reduces to a massive action for a
0-brane with charge $(q_1,q_2) = (-q^2,q^1) = T(1,p)$,
i.e. a $(1,p)$ 0-brane.

\section{Particle-String-Domain Wall Solutions in 9D}\label{9Dsugra}

As explained in the introduction, one can look for supergravity equivalents 
of the creation of strings in massive $D=9$ backgrounds. 
We first discuss the most general domain wall solutions of the 
9D gauged supergravities. Next, we construct a general solution of particles, 
strings and domain walls. We will show that it contains the
Kaluza-Klein reduction of the D0-F1-D8 solution \cite{Massar:1999sb} 
of Romans' supergravity as a special case.

\subsection{The Double Domain Wall in 9D}\label{9DdoubleDW}

We first consider domain walls in 9D gauged supergravities.
Each supergravity contains three mass parameters $m_1,m_2$ and $m_3$.
The supersymmetry transformations of the fermions, 
which we will need below, are given in \eqref{9Dsusy}.
We first note that one can always perform an $SL(2,\mathbb{R})$ 
transformation to set $m_1=0$.
This corresponds to diagonalising the symmetric version of the mass matrix
\be \label{diagmass}
M_{rs}=\epsilon_{rt}M^t{}_s=\tfrac{1}{2} \left(\begin{array}{cc}
\mu_1&0\\0&\mu_2
\end{array}\right)\, ,
\ee
where we have introduced 
\begin{align}
\mu_1=-m_2+m_3\,, \qquad
\mu_2=m_2+m_3\label{qs}\, .
\end{align}
By choosing appropriate values for $\mu_1$ and $\mu_2$ one can 
still cover each of the three conjugacy classes of
$SL(2,\mathbb{R})$, as shown in table \ref{massparameters}.

Using the projector ($\underline{y}$ indicates a tangent space direction, see
appendix A)
\begin{align}
  \Pi_{\text{DW}} = \tfrac{1}{2} (1+\gamma^{\underline{y}})\,,
\end{align}
one can find the 1/2 BPS domain wall solution 
\begin{align}
ds^2 &= (h_1 h_2)^{1/14} (-dt^2+dx_7^2)+(h_1 h_2)^{-3/7}dy^2\, ,\nn \\
\label{DW}
e^\phi&=h_1^{1/2} h_2^{-1/2}\,,\quad e^{\sqrt{7}\varphi}=(h_1 h_2)^{-1}\, , \\
\epsilon&=(h_1 h_2)^{1/56} \,\epsilon_0 \,, \nn
\end{align}
with harmonic functions
\begin{align}
  h_1=\mu_1 y+c_1 \,, \qquad
  h_2=\mu_2 y+c_2 \,.
\end{align}
This is the most general domain wall solution with $ISO(1,7)$ isometry and is in itself
a harmonic superposition of the domain walls with harmonics $h_1$ and $h_2$.
The two domain walls, which are parallel, thus form a threshold bound state.
Similar domain wall solutions to gauged supergravities have been found in $D=4, 5, 7$ and $8$ \cite{Bakas:1999fa, Bakas:1999ax, Alonso-Alberca:2003jq}.

The domain walls of the three 9D gauged supergravities were classified in 
\cite{Bergshoeff:2002mb} (for other discussions of 9D domain walls, see
\cite{Cowdall:2000sq,Nishino:2002zi}).
The results of \cite{Bergshoeff:2002mb} can be related to the general double domain wall \eqref{DW} by two operations. 
First of all one has to perform an $SL(2,\mathbb{R})$ transformation which chooses the frame with $\chi=0$. This can always be done without modifying the mass parameters. 
In addition one has to perform a coordinate transformation defined by
\begin{align}\label{cotr}
  h_1(y) h_2(y) = \tilde{H}(\tilde{y})^2 \,,
\end{align}
where the function $\tilde{H}(\tilde{y})$ appears in the metrics of 
\cite{Bergshoeff:2002mb} and is not necessarily harmonic.
Note that each different conjugacy class has a different function 
$\tilde{H}(\tilde{y})$ and therefore requires a different coordinate 
transformation.

The double domain wall can be truncated to a single domain wall
when restricting the constants $c_i$. The single domain walls
correspond to the situation where one of the domain walls has been taken away
or where the positions of the parallel domain walls
coincide. In table \ref{singleDW} we give the two possible truncations 
leading to single domain walls and the corresponding value of $\Delta$ as 
defined in \cite{Lu:1995cs}. Note that the $SO(2)$ case can not
be assigned a $\Delta$-value since it has vanishing potential, as 
already noted in 
\cite{Bergshoeff:2002mb}. The domain wall is carried by the 
non-vanishing massive 
contributions to the BPS equations. In other words, the potential is zero
but there is a non-vanishing superpotential.

\begin{table}[h]
\begin{center}
\begin{tabular}{||c||c|c|c|c||}
\hline \rule[-3mm]{0mm}{8mm}
  Gauge group & $(\mu_1,\mu_2)$ & $h_1$ & $h_2$ & $\Delta$ \\
\hline \hline \rule[-3mm]{0mm}{8mm}
  $\mathbb{R}$ & $(0,2m)$ & $c_1$ & $2m y + c$ & $4$ \\
\hline \rule[-3mm]{0mm}{8mm} 
  $SO(2)$ & $(2m,2m)$ & $2m y + c$ & $2m y + c$ & $\times$ \\
\hline
\end{tabular}
\caption{\it The single domain walls as truncations of the 
double domain wall solution. We give the two possible truncations and the corresponding value of $\Delta$. Note that there does not exist a $\Delta$-value in the $SO(2)$ case due to the vanishing of the potential.\label{singleDW}}
\end{center}
\end{table}

\subsection{Intersections with Particles and Strings}\label{9Dintersection}

We will start from a general Ansatz, respecting SO(7) symmetry. 
The fields are thus allowed to depend on $r=(x_1{}^2 + \ldots + x_7{}^2)^{1/2}$ and 
the transverse direction $y$. 
Our strategy will be to solve the BPS-equations obtained from the supersymmetry variations of the fermions. We will include the field strengths rather than the potentials in the Ansatz, since it is the former which enter the BPS equations. The independent fields and tensor components then are
\bea \label{fieldcontent}
  \{ g_{tt}, g_{ii}, g_{yy}, \phi, \varphi, \chi, F_{ty}, F_{ti}, F^r_{ty}, F^r_{ti}, H^r_{tyi}, H^r_{tij} \}\, ,
\eea
with the field strengths are given in terms of the potentials as
\be
F=dA\,,\quad F^r=dA^r-M^r{}_s B^s\,,\quad H^r=dB^r-A\wdg F^r\, .
\ee
Here $r=1,2$, the index $i$ is running from 1 to 7 and we have
dependence on $r$ and $y$ only. The upper indices on the curvatures
are the $SL(2,\RR)$-indices, which 9-dimensional supergravity inherits 
from type IIB in 10 dimensions.  Note that $F^r=\epsilon^{rs} F_s$ and the
same for $H^r$. We use a notation where $F_1 = - F^2$ is a NS-NS field
and $F_2 = F^1$ is a R-R field and similarly for $H$.
We have not included the four-form field strength $G$ nor tensor components lying purely in the spatial directions, since these correspond to $p$-branes with $p=2,\ldots,5$ which we do not want in our solution.
We use the following parametrisation for the Killing spinor:
\be \label{Killing}
\epsilon=f_1(r,y)\, e^{i f_2(r,y)}\,e^{i f_{\mu\nu}(r,y) \gamma^{\mu\nu}}\epsilon_0\, ,
\ee
with $f_1, f_2$ and $f_{\mu\nu}$ real.
In analogy with the solution \eqref{IIAsolution} for the Romans' mass parameter, we will assume that the $r$ and $y$ coordinates can be separated in a product, i.e. $f(y,r)=f(r) \, f(y)$.
This goes for all quantities in \eqref{fieldcontent} and \eqref{Killing} 
with the understanding that for the two dilatons we use $e^\phi$ and $e^\varphi$ rather than $\phi$ and $\varphi$.
This assumption will simplify the equations drastically.

The BPS-equations are obtained by requiring the spinor $\epsilon$ to be 
annihilated by the projection operators for the relevant branes. We search 
for solutions, which include domain walls, strings and particles.
Since we search for 1/4 BPS solutions the 3 projection operators corresponding
to the domain walls, strings and particles should not be independent.
In other words,  once we have two of the operators, the third should
follow as a combination of these. 
As mentioned in the previous section, we have the possibility of $SL(2,\RR)$ doublets of both the particles and the strings. By analysing the supersymmetry variations in Type IIB in $D=10$, it can be seen that the projectors for the F1- and the D1-strings are actually different, and this will therefore also be the case for the strings and particles in $D=9$. Choosing a specific string projector corresponds to choosing an $SL(2,\RR)$-frame. We take
the following projectors
\begin{align}
  \Pi_{\text{D0}} = \tfrac{1}{2} (1+\gamma^{\underline{0}}*) \,, \qquad
  \Pi_{\text{F1}} = \tfrac{1}{2} (1+\gamma^{\underline{0y}}*) \,, \qquad
  \Pi_{\text{DW}} = \tfrac{1}{2} (1+\gamma^{\underline{y}})\,,
\end{align}
where $*$ is seen as an operator, i.e. $*\epsilon=\epsilon^\ast$. Any third projector is implied by the other two: 
\bea
  \Pi_{\text{DW}} = \Pi_{\text{D0}} + \Pi_{\text{F1}} - 4 \, \Pi_{\text{D0}} \Pi_{\text{F1}}\, ,
\eea
and cyclic.
Since $\epsilon$ transforms under $SL(2,\RR)$, the choice of $SL(2,\RR)$-frame can be seen as a choice of $\epsilon$. To get the most general solution, we should keep the mass parameters as general as possible. We can, however, still perform $SL(2,\RR)$ transformations, which are upper triangular, without changing $\epsilon$. This can easily be seen by noting that $\epsilon$ transform as
\be
\epsilon\rightarrow\Big(\frac{c\tau^\ast+d}{c\tau+d}\Big)^{\frac{1}{4}}\epsilon
\ee
under the $SL(2,\RR)$-transformation
\be
\Lambda=\left(\begin{array}{cc}
a&b\\c&d
\end{array}\right)\, .
\ee
We see that $\epsilon$ is invariant for $c=0$.
The mass matrix transforms under $\Lambda$ as well. 
Even with $c=0$ we can always use $\Lambda$ to put $m_1$ to zero. 

Analysing the BPS equations we find that, in order to make up the
relevant projection operators, 
the following components must be put to zero:
\begin{align}
  & F_{ty}=F_{ti}=F^2_{ty}=F^2_{ti}=H^r_{tij}=H^1_{tyi}-\chi H^2_{tyi}=0
\label{hchi}\, .
\end{align}
The Bianchi identity for $F^2$ reads $dF^2=\tfrac{1}{2} \mu_1 H^1\label{bi}$.
Since $F^2=0$, this will lead to further restrictions when $\mu_1$ is 
non-vanishing.
We find that $H^1=0$ and, using (\ref{hchi}), also $\chi H^2_{tyi} = 0$.
We require $H^2 = - H_1 \ne 0$, since otherwise no F-strings would be present
and we conclude that $\chi = 0$ if $\mu_1 \ne 0$. 
If $\mu_1=0$, one can draw the same conclusion but from a different argument.
In this case, the BPS equations directly imply $\partial_\mu \chi = 0$ and 
therefore $\chi$ is a constant.
The only non-zero mass parameter $\mu_2$ gauges the $\mathbb{R}$ subgroup of $SL(2,\mathbb{R})$, which shifts the axion.
Thus one can always use a global gauge transformation to set $\chi = 0$.
Then (\ref{hchi}) implies $H^1=0$. 
On top of this we take $F^1_{ty}=0$ since a non-zero value requires 
D0-brane sources smeared on the domain-wall worldvolume 
and we want to avoid such 'walls' of D0-branes (see also the introduction 
and Conclusions). 
We thus find that, for all values of the mass parameters, we are 
left with just two non-vanishing tensor components, 
$F^1_{tr}$ and $H^2_{tyr}$. 

We now substitute our Ansatz in the supersymmetry variations of the fermions,
which are given in appendix \ref{reductionansatze}.
Requiring $\delta\lambda = \delta\tilde{\lambda} = 0$ leads to
first order differential equations for the dilatons, which 
also depend on the other fields. From $\delta\psi_\mu = 0$ we get first 
order differential equations for the metric as well as differential equations 
for the functions which define $\epsilon$. 
At this stage, the solution to the BPS-equations contains two undetermined 
functions,  one depending on $r$ and one depending on $y$. 
The latter can be fixed arbitrarily by using a general coordinate 
transformations in $y$. To determine the function of $r$, 
we need at least one field equation, e.g. the one for $\varphi$. 
We have computed this field equation, and the result is that the 
$r$-dependent function can be expressed in terms of a harmonic function. 
The resulting particle-string-domain wall solution
can be expressed in a unified way, i.e. including all cases 
$\alpha^2= 0$, $\alpha^2 > 0$ and $\alpha^2 <0$, as follows
\begin{align}\nn \label{9Dsolution}
ds^2&=-(h_1 h_2)^{1/14} h^{-11/7}dt^2+(h_1 h_2)^{-3/7} h^{-4/7} dy^2+(h_1 h_2)^{1/14} h^{3/7}dx_7^2\, ,\\\nn
e^\phi&=h_1^{1/2} h_2^{-1/2}\,,\quad e^{\sqrt{7}\varphi}=(h_1 h_2)^{-1} h\, ,\\
A^1_t&=-h_2^{1/2} h^{-1}\,,\quad B^2_{ty}=h_2^{-1/2} h^{-1}\, ,\\\nn
\epsilon&=(h_1 h_2)^{1/56} h^{-11/28}\,\epsilon_0\, .
\end{align}
The solution is given in terms of three harmonic functions
\begin{align}
h_1=\mu_1 y+c_1 \,, \qquad
h_2=\mu_2 y+c_2 \,, \qquad
h=1+\frac{Q}{r^5} \,.
\end{align}
The $\mu$'s are given in terms of the mass parameters in (\ref{qs}) and 
$c_1$ and $c_2$ are integration constants.
Just as in $D=10$, the solution is a harmonic superposition of D-particles, F-strings and domain walls with string and particle fluxes
\begin{align}
  \mathcal{Q}_1 & = e^{-\phi-\varphi/\sqrt{7}} \star (d B^2) = Q \, h_2{}^{1/2} \, d \Omega_6 \,, \notag \\
  \mathcal{Q}_0 & = - e^{\phi + 3 \varphi/\sqrt{7}} \star (d A^1 - \tfrac{1}{2} \mu_2 B^2) = Q \, h_2{}^{-1/2} \, dy \wedge d \Omega_6 \,,
\label{D0F1}
\end{align}
with $d\Omega_6$ the volume form of the $S^6$. 
The charges are obtained by integrating the fluxes over 
$S^6$ and $S^6 \times \mathbb{R}$, respectively, where the $S^6$,
together with the 7D radius $r$, 
spans $\mathbb{R}^7$ and $\mathbb{R}$ covers the $y$-range.
The flux distributions are related by 
\begin{align}
  d \mathcal{Q}_1 = - \tfrac{1}{2} \mu_2 \mathcal{Q}_0 \,,
\end{align}
as required by the $B^2$ equation of motion.

Of course one can perform an $SL(2,\mathbb{R})$ transformation on the 
solution \eqref{9Dsolution} and obtain intersections with more general strings and particles.
The $SL(2,\mathbb{R})$ generalised flux distributions are given by
\begin{align}
  \mathcal{Q}_1{}^r & = e^{-\varphi/\sqrt{7}} \mathcal{M}^{r}{}_{s} \star (d B^s) 
    = q_1{}^r \mathcal{Q}_1 \,, \notag \\
  \mathcal{Q}_0{}^r & = e^{3 \varphi/\sqrt{7}} \mathcal{M}^{r}{}_{s} \star (d A^s - M^s{}_t B^t) 
    = q_0{}^r \mathcal{Q}_0 \,.
\end{align}
In this notation the F-strings and D-particles \eqref{D0F1} have charges $q_1{}^r=(1,0)$ and $q_0{}^r=(0,1)$.
A transformation with parameter
\begin{align}\label{sl2r}
  \Lambda = \left(\begin{array}{cc} r & p \\ s & q \end{array}\right) \in SL(2,\mathbb{R}) \,, 
\end{align}
would take the distributions of F-strings and D-particles \eqref{D0F1} to $q_1{}^r=(r,s)$ and $q_0{}^r=(p,q)$. This
corresponds to $(p,q)$-particles and $(r,s)$-strings subject to the condition
$qr - ps = 1$.
Furthermore, the $SL(2,\mathbb{R})$ transformation (\ref{sl2r})
rotates the diagonal background \eqref{diagmass} into
\begin{align}
  M_{rs} = \tfrac{1}{2} \left(\begin{array}{cc} -m'_2 + m'_3 & m'_1 \\
   m'_1  & m'_2 + m'_3 \end{array}\right) = 
  \tfrac{1}{2} \left(\begin{array}{cc} q^2 \mu_1 + s^2 \mu_2 & -pq \mu_1 - rs \mu_2 \\
    -pq \mu_1 - rs \mu_2 & p^2 \mu_1 + r^2 \mu_2 \end{array}\right) \,.
\label{mass}
\end{align}
From now on we will omit the primes on the mass parameters.
Thus we find that the most general intersection of $(p,q)$-particles, $(r,s)$-strings and an $(m_1,m_2,m_3)$-domain wall is subject to two conditions:
\begin{itemize}
\item The $SL(2,\mathbb{R})$ condition $qr - ps = 1$ should be
satisfied. This condition requires orthogonality of the strings and particle charges. It can be expressed as $q_1{}^r q_0{}_r = 1$.
\item The form of the mass matrix \eqref{mass} is given in 
(\ref{mass}). This mass matrix contains only two independent 
parameters $\mu_1$ and $\mu_2$ rather than three for an arbitrary but symmetric mass matrix. This restriction corresponds to $q_1{}^r M_{rs} q_0{}^s = 0$.
\end{itemize}
The two orthogonality conditions are manifestly $SL(2,\mathbb{R})$-invariant and the parameters $\mu_1$ and $\mu_2$ specify the only $SL(2,\mathbb{R})$-orbits that solves the BPS equations.
 
\begin{figure}[h]
  \centerline{\epsfig{file=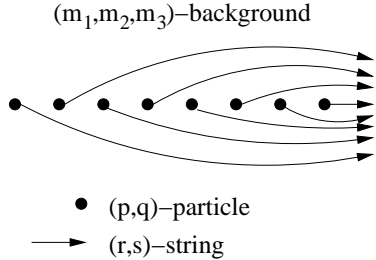,width=.35\textwidth}}
  \caption{\it The creation of strings in 9D: a (continuous) distribution 
of $(p,q)$-particles with
  a monotonically increasing distribution of emanating oriented (r,s)-strings  in an $(m_1,m_2,m_3)$-background. There are two orthogonality
 conditions on the charges. Note that no source term has been included.} \label{HWpq}
\end{figure}

The physical picture consists of a distribution of particles from which strings are emanating towards the domain wall, like in the IIA case. 
However, we now have an $SL(2,\mathbb{R})$ generalisation of $(r,s)$-strings stretching between $(p,q)$-particles in an $(m_1,m_2,m_3)$-background with two orthogonality conditions.
The two conditions reduce the seven parameters to five, three of which correspond to the $SL(2,\mathbb{R})$ freedom while the two remaining parameters are $\mu_1$ and $\mu_2$.
In addition the charge $Q$ is the unit string charge.
The general solution is illustrated in figure \ref{HWpq}.
The interval in this case is Max$(-\mu_1/c_1, -\mu_2/c_2) < y < 0$ with all 
$\mu_i$ and $c_i$ positive.
Note that the charge distribution of the strings is not linear, as opposed to
 the massive IIA solution in 10D.
This is due to the freedom of reparametrising the $y$-coordinate; 
the important feature is that $\mathcal{Q}_0$ is continuous and positive, 
implying $\mathcal{Q}_1$ to be monotonically increasing on this interval.
In the discussion we will comment on the possibility of extending the 
interval to a globally well-defined solution.

One can take different limits of the general solution (\ref{9Dsolution}).
First of all, one can set the parameter $\mu_1=0$.
This case corresponds to the reduction of the massive IIA solution of 
\cite{Massar:1999sb}
and indeed the Kaluza-Klein reduction of \eqref{IIAsolution} along
one of the worldvolume directions of the D8-brane
gives (changing $y$ to $\tilde{y}$ for reasons that will become clear shortly)
\begin{align}\nn
ds^2 &= -H^{1/7}h^{-11/7}dt^2+H^{8/7}h^{-4/7}d\tilde{y}^2+
H^{1/7}h^{3/7}dx_7^2\, ,\\
e^\phi&=H^{-1}\,,\quad e^{\sqrt{7}\varphi}=H^{-2}h\, ,\\\nn
B^{2}_{t\tilde{y}}&=h^{-1}\,,\quad A^{1}_t=H h^{-1}\, ,
\end{align}
where the harmonic functions are defined as
\be
H=c+m\tilde{y}\,,\quad h=1+\frac{Q}{r^5}\, .
\ee
The above is a special solution to the 9-dimensional gauged supergravity 
where the mass parameters obey $\mu_1 = 0$ and $\mu_2 = 2 m$.
Exactly the same identifications were found
in the case of the reduced massive D-particle, see eq.~\eqref{mr}. 
It is related to the generic solution \eqref{9Dsolution} by a coordinate transformation $y=y(\tilde{y})$ defined by $h_2(y) = H(y)^2$  which is a
special case of \eqref{cotr}.

Another possible truncation of the general solution
\eqref{9Dsolution} is obtained by setting
both mass parameters $\mu_1$ and $\mu_2$ equal to zero.
This yields a harmonic superposition of the F-string solution with a distribution of D-particles on it.
The two charge distributions are related (both are linear in $Q$) and therefore it is impossible to obtain either one separately.

\section{Discussion and Conclusions}\label{discussion}

In this paper we have investigated $(p,q)$-particles in gauged 9D supergravities from two
perspectives: the $(p,q)$-particle brane action in a massive background and a corresponding
supergravity solution.
It is interesting that in both approaches it is possible to generalise the striking
features of the corresponding 10D IIA system:
\begin{itemize}
\item
In the $(p,q)$-particle brane action in a massive 9D background, see \eqref{particleaction}, 
a mass term similar to the coupling of the D-particle Born-Infeld vector to the background 
\eqref{term1} occurs.
\item
The supergravity solution \eqref{9Dsolution} consists of a smeared distribution of $(p,q)$-particles, 
from which $(r,s)$-strings are emanating and ending on the $(m_1,m_2,m_3)$-domain wall. There
are two orthogonality conditions on the seven parameters, as discussed in subsection 
\ref{9Dintersection}.
This is the natural generalisation of the 10D IIA solution \eqref{IIAsolution}.
\end{itemize}
We now would like to discuss a number of open issues associated to the new particle-string-domain wall system.

At first sight the mass term in the $(p,q)$-particle brane action seems non-sensical since 
it introduces non-propagating worldline vectors $V_r$ as Lagrange multipliers.
The corresponding constraints seemingly imply that the mass parameters must be vanishing, which would lead to the massless background.
The situation is identical in 10D IIA, where the na\"{\i}ve field equation for the Born-Infeld vector is $m=0$. As discussed in the introduction and explained in \cite{Banks:1997zs,Bergman:1997gf,Bachas:1998kn}, the resolution lies in the inclusion of extra degrees of freedom corresponding to strings stretched between the particle and the domain wall. We suggest that a similar mechanism occurs in 9D and relates the mass term to the creation of strings. It would be interesting to see whether this could reproduce the orthogonality constraints on the charges found in our supergravity solution.

As discussed in the introduction, the occurrence of domain walls with positive tension leads to 
a harmonic function that vanishes at a point in the transverse space. 
To avoid this, one has to include orientifold planes with 
negative tension as well, which can be introduced by modding out the theory with a 
$\mathbb{Z}_2$-transformation. In 10D IIA the relevant symmetry is $I_y \Omega$, which introduces
(in the case of $y$ compact) 16 D8-branes and their images and two O8-planes.
In 9D the relevant $\mathbb{Z}_2$-symmetry can be obtained from the IIA transformation $I_y \Omega$
by the reduction in a direction other than $y$.
Alternatively, one could reduce the IIB transformation $(-)^{F_L} I_{xy} \Omega$ in the $x$ direction.
Upon reduction these give the same transformation and therefore are T-dual \cite{Bergshoeff:2001pv}.
In particular, the 9D $\mathbb{Z}_2$-symmetry acts on the mass parameters as $M_{rs} \rightarrow - M_{rs}$.
Thus all three mass parameters flip sign. However, one can always use an $SL(2,\mathbb{R})$-transformation to set $m_1=0$.
Then one is left with $\mu_1$ and $\mu_2$ and since both mass parameters flip sign, one introduces orientifold planes which carry a charge of $-16$ with respect to both $\mu_1$ and $\mu_2$.
Taking $y$ compact (for a non-compact transverse space the discussion is analogous), one also has to introduce a number of positive tension branes to cancel 
the total charges.
For the $\mu_2$-charge this correspond to 32 D7-branes.
The cancellation of $\mu_1$-charge requires 32 Q7-branes, which are defined as S-duals of the 
D7-branes.
Thus the following picture seems to emerge:
\begin{itemize}
\item
Two orientifold planes, one at each of the fixed points of the $S^1$, each carrying a charge of $(-16,-16)$ with respect to the two mass parameters $(\mu_1,\mu_2)$.
\item
Sixteen D7-branes and their images, located at arbitrary points between the two O7-planes and each carrying a charge of $(0,1)$.
\item
Sixteen Q7-branes and their images, defined as S-duals of the D7-branes, also distributed between the two O7-planes and each carrying a charge of $(1,0)$.
\end{itemize}
Depending on the positioning of the various 7-branes, the mass parameters can take different values.
Note that the gauge group can change when passing through a 7-brane, since it can affect only $\mu_1$ or $\mu_2$ and thus $\alpha^2 = - \tfrac{1}{4} \mu_1 \mu_2$ need not be invariant.
The reduction of the type I${}^\prime$ theory would correspond to a special case of this general set-up, 
in which eight of the Q7-branes and their images are positioned at each O7-plane, thereby
cancelling the $(-16,0)$ charge and inducing $\mu_1=0$ everywhere in the bulk\footnote{
  Toroidal compactifications of type I${}^\prime$ string theory have been considered in 
  \cite{Chaudhuri:2000aa} from a somewhat different point of view.
  It would be interesting to link its results to the analysis of this paper.
  We thank S.~Chaudhuri for a discussion on this point.
}.

\begin{figure}[h]
  \centerline{\epsfig{file=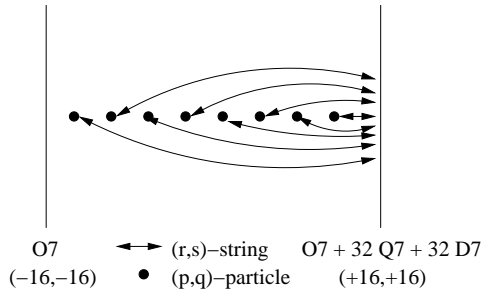,width=.45\textwidth}}
  \caption{\it The creation of strings in 9D on $S^1 /\mathbb{Z}_2$: a (continuous) distribution of 
  (p,q)-particles with a monotonically increasing distribution of unoriented (r,s)-strings 
  ending on the D7- and Q7-branes. } \label{HWDQO}
\end{figure}

The 9D particle-string-domain wall solution \eqref{9Dsolution} corresponds to a region between
two domain walls on the $S^1 /\mathbb{Z}_2$, as illustrated in figure \ref{HWpq}. 
We now discuss the possibility of extending this to a globally well-defined solution by including source terms for the domain walls and the particle-string
intersection. Note that all tensor components of \eqref{9Dsolution}
are even under the relevant 9D $\mathbb{Z}_2$-symmetry. In fact, the reason to discard the
possibility of non-zero $F^1_{ty}$ was its odd transformation under this $\mathbb{Z}_2$-symmetry\footnote{This is similar to the IIA discussion in the 
introduction concerning the solution of \cite{Janssen:1999sa}.}.
Thus one is led to think that it is possible to embed the solution \eqref{9Dsolution} in a
globally well-defined solution on $S^1 /\mathbb{Z}_2$. This is illustrated in figure \ref{HWDQO}, where all D7- and Q7-branes are taken to coincide with one of the orientifold planes.
It would be interesting to investigate the boundary conditions in a manner analogous to the IIA analysis of \cite{Kallosh:2001zc}.

To conclude, the results of this paper suggest new possibilities of string creation in nine dimensions that are not the result of the reduced Type I${}^\prime$ mechanism. 
It would be a challenge to confirm this phenomenon from a string theory analysis.

\section*{Acknowledgements}

\noindent 
We thank Fabio Riccioni, who was involved at an early stage of this work. 
This work is supported in part by the European Community's Human
Potential Programme under contract HPRN-CT-2000-00131 Quantum
Spacetime, in which the University of Groningen is associated with the
University of Utrecht.  The work of U.G. is part of the
research program of the ``Stichting voor Fundamenteel Onderzoek der
Materie'' (FOM).

\appendix

\section{Conventions}\label{conventions}

Greek indices $\mu,\nu,\rho,\ldots$ denote curved spacetime indices while 
Latin $a,b,c\ldots$ indices are tangent spacetime indices. They are related by the
vielbeins $e_{\mu}{}^{a}$ and inverse vielbeins $e_{a}{}^{\mu}$. Explicit
indices $0, \ldots, 9$ are underlined for tangent spacetime coordinates and not underlined
for the curved coordinates. We use hats for 10-dimensional objects (both IIA and IIB) 
and no hats for 9-dimensional objects. We use mostly plus signature $(-+\cdots+)$ and
our metric is always given in Einstein frame.
$i,j,\ldots$ are worldvolume indices while $r,s,\ldots$ are $SL(2,\RR)$-indices. 
Indices of the latter type can be raised and lowered with $\epsilon^{rs}$ and $\epsilon_{rs}$, with $\epsilon^{12}=+1$ and $\epsilon_{12}=-1$.

\section{Reduction Ans\"atze and 9D Supersymmetry}\label{reductionansatze}

For type IIB, the reduction is
\begin{align}\nn
\hat{e}_{\hat{\mu}}{}^{\hat{a}}&=\left(\begin{array}{cc}e^{\sqrt{7}\varphi/28}e_\mu{}^a&-e^{-\sqrt{7}\varphi/4}A_\mu\\0&e^{-\sqrt{7}\varphi/4}\end{array}\right)\, ,\\
(\hat{B}_r)_{\mu\nu}&=(\Omega^{-1T})_r{}^s(B_s)_{\mu\nu}\, ,\\\nn
(\hat{B}_r)_{\mu x}&=-(\Omega^{-1T})_r{}^s(A_s)_\mu\, ,\\\nn
\hat{{\cal M}}^{rs}&=\Omega^r{}_p \, \Omega^s{}_q \,{\cal M}^{pq}\, .
\end{align}
For type IIA, the reduction is
\begin{align}\nn
\hat{e}_{\hat{\mu}}{}^{\hat{a}}&=\left(\begin{array}{cc}e^{\phi/16-3\varphi/16\sqrt{7}}e_\mu{}^a&e^{-7\phi/16+3\sqrt{7}\varphi/16}(A_1)_\mu\\0&e^{-7\phi/16+3\sqrt{7}\varphi/16}\end{array}\right)\, ,\\\nn
\hat{B}_{\mu\nu}&=-(B_1)_{\mu\nu}+2(A_1)_{[\mu}A_{\nu]}\, ,\\
\hat{B}_{\mu x}&=-A_\mu\, ,\\\nn
\hat{A}_\mu&=-(A_2)_\mu+\chi(A_1)_\mu\, ,\\\nn
\hat{A}_x&=-\chi\, ,\\\nn
\hat{\phi}&=\frac{3\phi}{4}+\frac{\sqrt{7}\varphi}{4}\, .
\end{align}

The supersymmetry variations of the fermions in 9D gauged supergravity are given by 
(in the conventions of \cite{Bergshoeff:2002nv})
\begin{align}
\delta \psi_\mu
&= D_\mu \epsilon + \ft{i}{16} e^{-2 \varphi / \sqrt7} \left( \ft57 \gamma_{\mu} \gamma^{(2)} - \gamma^{(2)} \gamma_{\mu} \right) F_{(2)} \epsilon\, , \nn\\
& \quad - \ft{1}{8\cdot 2!} e^{3 \varphi/ {2\sqrt7}} \left( \ft57 \gamma_{\mu}\gamma^{(2)} - \gamma^{(2)}\gamma_{\mu} \right) \,  e^{\phi/2} \left( F^{1} - \tau F^{2} \right)_{(2)} \epsilon^*\, , \notag \nn\\
& \quad + \ft{i}{8\cdot 3!}e^{- \varphi / {2 \sqrt7}} \left( \ft37 \gamma_{\mu}\gamma^{(3)} + \gamma^{(3)}\gamma_\mu \right)  e^{\phi/2} \left( H^{1} - \tau H^{2} \right)_{(3)}\epsilon^*\, , \notag \nn\\
& \quad - \ft{1}{8\cdot 4!}e^{\varphi / \sqrt7} \left( \ft{1}{7} \gamma_{\mu} \gamma^{(4)} - \gamma^{(4)} \gamma_{\mu} \right) G_4 \epsilon +\ft{1}{7} \gamma_{\mu} W \epsilon \displaybreak[2] \,,\nn\\[2mm]
\delta \tilde\lambda & = i \slashed \partial \varphi \, \epsilon^* - \ft{1}{\sqrt{7}} e^{-2 \varphi/ \sqrt7} \slashed F \epsilon^* - \ft{3i}{2\cdot 2!\sqrt{7}}e^{3\varphi/{2\sqrt7}} e^{\phi/2}\gamma^{(2)}\left( F^{1} - \tau^* F^{2} \right)_{(2)} \epsilon\, , \notag \\
& \quad + \ft{1}{2\cdot 3!\sqrt{7}}e^{-\varphi/ {2 \sqrt7}}e^{\phi/2}\gamma^{(3)}\left(H^{1} - \tau^* H^{2} \right)_{(3)}\epsilon\, , \nn\\
& \quad + \ft{i}{4!\sqrt{7}} e^{\varphi/ \sqrt7} \slashed G_4 \epsilon^* +4 i\, \frac{\delta W}{\delta \varphi}\, \epsilon^*\, ,\nn\\[2mm]
\delta \lambda
&= i \slashed{\partial} \phi \, \epsilon^* - e^\phi \slashed{\partial} \chi \, \epsilon^* - \ft{i}{2\cdot 2!} e^{3\sqrt7\varphi/14} e^{\phi/2}\gamma^{(2)}\left( F^{1} - \tau F^{2} \right)_{(2)}\epsilon\, , \nn\\
& \quad - \ft{1}{2\cdot 3!}e^{-\sqrt{7}\varphi/14}e^{\phi/2}\gamma^{(3)}\left(H^{1} - \tau H^{2} \right)_{(3)}\epsilon +4i\, \Big(\frac{\delta W}{\delta \phi} 
      +i e^{- \phi} \frac{\delta W}{\delta \chi}\Big)\,\epsilon^* \, ,
\label{9Dsusy}
\end{align}
where the superpotential is given in terms of the scalars and mass parameters as
\begin{align}
  W = \ft14 e^{2\varphi/\sqrt{7}}
    \left( m_2 \sinh(\phi) + m_3 \cosh(\phi) + m_1 e^\phi\chi
    - \ft12 (m_2-m_3)e^\phi \chi^2\right ) \, .
\end{align}

\section{Formulations of the $(p,q)$-String Action}\label{pqformulations}

In this appendix we will show how to relate the $SL(2,\mathbb{R})$-covariant form
of the $(p,q)$-string action \eqref{l} to other formulations of the D- or F-string
\cite{Cederwall:1997ts, Schmidhuber:1996fy}.
The basic idea is that a vector on a worldsheet carries no degrees of freedom and
therefore can be dualised into a constant. Thus there are different ways to represent
the same field equations.

To bring the action into the standard form, it is convenient
to add a surface term, which doesn't change the equations of motion
\be\label{ss}
S=\h\,\int d^2\sigma\,\lambda\,(\hat{g}+\hat{{\cal M}}^{rs}\ast \hat{F}_r 
\ast \hat{F}_s)+\hat{q}^r\int d^2\sigma\,\ast d\hat{V}_r\, .
\ee
Before adding the surface term, the field equation for $\hat{V}_r$ says 
that $\lambda\,
\hat{{\cal M}}^{rs} \ast \hat{F}_s$ is a constant $c^r$. After adding the 
surface 
term, one can effectively regard $d\hat{V}_r$ as an independent field and this 
identifies the constant $c^r$ with $c^r = -\hat{q}^r$. Using the $\lambda$ and 
$d\hat{V}_r$ equations of motion one obtains in the presence of the
surface term the following equivalent set of equations:
\bea
&& \hat{g}=-\hat{{\cal M}}^{rs}\ast \hat{F}_r \ast \hat{F}_s\, , \qquad
\lambda\,\hat{{\cal M}}^{rs} \ast \hat{F}_s=-\hat{q}^r\, .
\eea
The latter equation has the solution
\be
\lambda \Phi= i \, e^{\hat{\phi}/2} ( \hat{\tau} \hat{q}_1 + \hat{q}_2) \, ,
\ee
with $\hat{q}_r=\epsilon_{rs}\hat{q}^s$ and $\hat{\mathcal{M}}^{rs}$ and $\hat{\tau}$ defined in subsection \ref{pqstring}.
After integrating out both vector fields, the action takes the form
\be\label{sqrtS}
  S=\int d^2\sigma\,\sqrt{\hat{q}_r \hat{\mathcal{M}}^{rs} \hat{q}_s}
  \sqrt{-\hat{g}}+\hat{q}^r\int d^2\sigma \ast \hat{B}_r\, .
\ee
Since the string charges are quantised, we can write $\hat{q}^r=T\,\hat{n}^r$, 
where $T$ is the string tension and $n^r$ are integers.
For pure NS-NS charge, $\hat {n}^r = (1,0)$, this action reduces to the 
standard form (using Einstein frame)
\be\label{standard}
S=T\int d^2\sigma\, \sqrt{-\hat{g}_s}+
T\int d^2\sigma \ast \hat{B}\, ,
\ee
where $\hat{g}_s$ is the metric in the string frame and
 $\hat{B} \equiv \hat{B}_1$. 

Instead of integrating out both vectors we can also integrate out only 
$\hat{V}_2$ and end up with the D-string action. To get the DBI-form, 
we need to rewrite the action by including the physical scalars via 
(\ref{scal})
\begin{align}\nn
S&=\h \int\,d^2\sigma\,\lambda\Big({\rm det}\hat{g}+e^{-\hat{\phi}}
(\ast \hat{F}_1)^2+e^{\hat{\phi}}(\ast \hat{F}_2+\hat{\chi}\ast\hat{F}_1)^2
\Big)+\hat{q}^r\int\,d^2\sigma\ast d\hat{V}_r\, ,\\
&=\h\int\,d^2\sigma\,\tilde{\lambda}\Big({\rm det}(\hat{g}_s+\hat{F}_1)+
(\ast\hat{F}_2+\hat{\chi}\ast{F}_1)^2\Big)+\hat{q}^r\int\,d^2\sigma\,\ast 
d\hat{V}_r\, ,\label{dbi2}
\end{align} 
where $g_s$ is the metric in the string frame, $\tilde{\lambda}=
\lambda e^{\hat{\phi}}$ and we have used that $(\ast\hat{F}_1)^2=
((\hat{F}_1)_{01})^2={\rm det}\hat{F}_1$. Integrating out $\tilde{\lambda}$ 
and $\hat{V}_2$ and putting $(\hat{q}^1,\hat{q}^2)=(0,-T)$, i.e. 
$(\hat{q}_1,\hat{q}_2)=(T,0)$, yields
\be
S=T\int\,d^2\sigma\,e^{-\hat{\phi}}\sqrt{-{\rm det}(\hat{g}_s+\hat{F}_1)}-
T\int\,d^2\sigma\,(\ast \hat{B}_2-\hat{\chi}\ast\hat{F}_1)\, ,
\ee
which is the ordinary D-string action.

\bibliography{maction}
\bibliographystyle{toine}

\end{document}